# Integrated Hardware Annealing based on Langevin Dynamics for Ising Machines

Yongchao Liu, Lianlong Sun, Michael Huang, Hui Wu
Department of Electrical and Computer Engineering, University of Rochester
{yongchao.liu, lianlongsun, michael.huang, hui.wu}@rochester.edu

***Abstract*—Ising machines are non-von Neumann machines designed to solve combinatorial optimization problems (COP) by searching for the ground state, or the lowest energy configuration, within the Ising model. However, Ising machines often face the challenges of getting trapped in local minima due to the complex energy landscapes. Hardware annealing algorithms help mitigate this issue by using a probabilistic approach to steer the system toward the ground state. In this paper, we present a hardware annealing algorithm for Ising machines based on Langevin dynamics, a stochastic perturbation by random noise. Theoretical analysis, system-level design, and detailed circuit design are carried out. We evaluate the performance of the algorithm through chip-level simulation using a standard 65-nm CMOS technology to demonstrate the algorithm's efficacy. The results show that the proposed hardware annealing algorithm effectively guides the system to reach the ground state with a probability of 86.5%, significantly improving the solution quality by 97.5%. Further, we compare the algorithm with state-of-the-art hardware annealing methods through behavioral-level simulations, highlighting its improved solution quality alongside a 50% reduction in time-to-solution.**

***Index Terms*—Combinatorial optimization problems, Ising machine, Langevin dynamic, Random noise, Annealing**

## I. Introduction

The advancement in modern technologies, such as semiconductor chip design, artificial intelligence, medicine, and applied mathematics, has led to the emergence of crucial challenges that must be addressed, wherein solving combinatorial optimization problems (COPs) that aim to find the optimal solution among discrete solutions becomes rather important [1] - [3]. A large fraction of COPs are classified as non-deterministic polynomial-time (NP-hard) problems, in which the solution space expands exponentially with the number of variables involved. Traditional computing paradigms encounter considerable obstacles due to the exponential growth in solution time and resource demands.

Ising machine has been studied as a specialized non-von Neumann machine capable of searching for the low-energy state of the Ising model. With COPs mapped onto the Ising machine, it can look for the optimal solution of NP-hard COPs effectively, which substantially reduces computation time and requirements of hardware resources, compared to the exhaustive search on conventional computing devices. Recently, CMOS-based Ising machines, such as digital and analog annealers, have gained considerable attention due to their high solution quality and low power consumption [4]. However, these machines cannot always reach the ground state, due to the possibility of getting trapped in local minima. It can significantly degrade the solution quality, particularly as the number of spins increases and the solution space expands. Hence, annealing algorithms within the Ising machine need to be studied to help escape from the local minima.

Various annealing techniques, especially the simulated annealing based annealing algorithm [5] - [7], have been extensively explored for CMOS-based Ising machines. These algorithms rely on spin-flip operations to allow the system to escape local minima, which involves storing the current spin configurations, detecting their values, and randomly flipping a subset of spins through digital control mechanisms. Thus, these techniques are widely adopted in digital annealers, where the entire annealing process is efficiently managed. In contrast, analog annealers have the ability to seek low-energy states spontaneously once COPs are mapped onto them and do not require constant digital controls. Therefore, the requirement for spin-flip operations in analog annealers presents a challenge, as it complicates the annealing process. Additionally, enabling spin-flips requires multiple memory accesses and spin-swapping operations to update a subset of spins. This introduces significant time-to-solution (TTS) overhead, making it a bottleneck in terms of efficiency.

To address this issue, we proposed a physical hardware annealing algorithm based on Langevin dynamics, a stochastic process with perturbation by random noise. The annealing algorithm is evaluated to effectively solve all-to-all connected COPs both on the chip level and at the behavior level, offering a high likelihood of reaching the ground state and significant improvement of solution quality within a short TTS. The rest of the paper is organized as follows: Section II illustrates the relevant background. Section III presents a theoretical analysis. Section IV describes the hardware implementations of the proposed annealing schedule, including system-level and detailed circuit design. Section V evaluates the proposed annealing algorithm. Section VI concludes the paper.

## II. Background

### A. Ising model

Ising model [8] is a statistical model used to describe the energy of a many-body system composed of discrete and coupled spins, as shown in Fig. 1(a). The Ising model can be applied to physical systems in which the spins($\sigma$) take on binary values of either +1 or -1, and interact with each other through the coupling coefficient denoted by $J_{ij}$. The external energy interaction $h_i$ is an energy perturbation by the external magnetic field $\mu$ to individual spin. In this system, the spins automatically flip to align themselves according to the state of their coupled spins in an effort to reduce the system's energy. The Ising Hamiltonian provides a mathematical function to describe the system energy as in Eqn. (1):

$$E = -\sum_{(i<j)} J_{ij}\sigma_i\sigma_j - \mu\sum_i h_i\sigma_i \tag{1}$$

A physical system with such an Ising Hamiltonian has a natural tendency to decrease its energy and reach the equilibrium, depicted in Fig. 1(b), as if solving optimization problems through energy computation with Eq. 1 as the objective function. Therefore, by engineering a physical system with programmable $J_{ij}$ and $h_i$, it can function as a specialized energy-based computer capable of solving COPs.

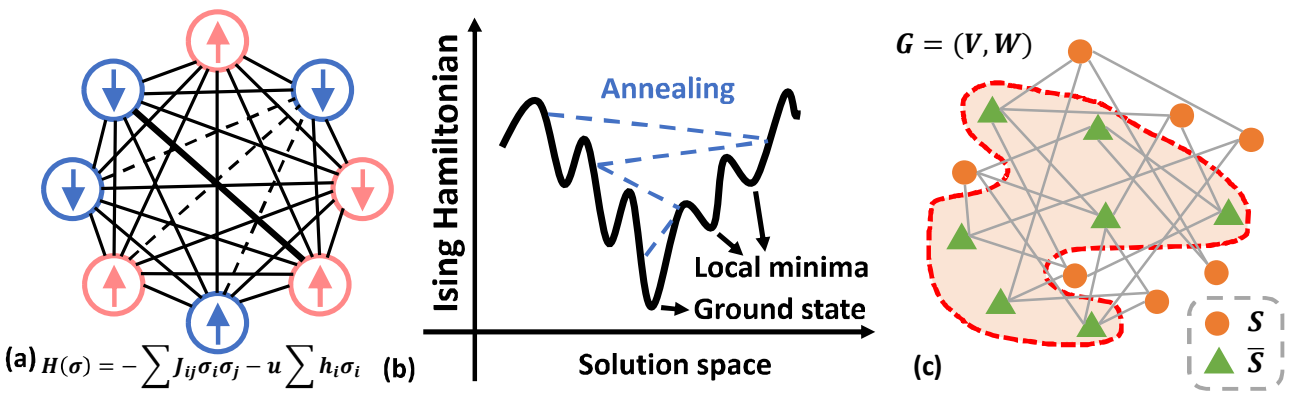

Fig. 1. (a) Ising model (b) Ising Hamiltonian navigation (c) Max-cut problem.

In this paper, we adopt the *Bistable Resistive-coupled Ising Machine* (BRIM) [3] as the Ising machine substrate to evaluate the performance of the annealing algorithm. The BRIM module shown in Fig. 2 describes the high-level architecture of the Ising machine. In this figure, $V_i$ represents the $i^{th}$ spin with a capacitor in it for current storage and is coupled to other spins via programmable resistive couplers $J_{ij}$. The coupling coefficient is stored in the memory and will be used to program the couplers via DACs.

### B. Langevin dynamics

The Langevin dynamics was initially formulated to model the dynamics of molecular systems influenced by random noise, with Brownian motion being the most common example. Considering particles with mass $m$ doing Brownian motion in a potential of particles $U(\boldsymbol{x})$, the Langevin dynamics of the system can be expressed as [10]:

$$m\ddot{\boldsymbol{x}} = -\beta\dot{\boldsymbol{x}} - \nabla U(\boldsymbol{x}) + \sqrt{2\beta kT}\cdot\boldsymbol{\xi}(t) \tag{2}$$

where $\boldsymbol{x}^\top = (x_1, x_2 \cdots x_N)^\top$ represents the position of particles. $k$ is the Boltzmann constant and $T$ is temperature. $\beta$ is the friction coefficient of the fluid. $-\beta\dot{\boldsymbol{x}}$ represents the drag force acting on the particles, gradually decreasing the initial velocity until zero over a sufficiently long period. $\boldsymbol{\xi}(t)$ denotes the random impact (e.g., from bombardment of surrounding molecules) with properties of $\langle\xi_i(t)\rangle = 0$ and $\langle\xi_i(t)\xi_i(t')\rangle = C\delta(t-t')$, meaning the random forces are zero mean and unrelated random shock, which align with the properties of white noise.

The evolution of the joint probability density function $P(x,t)$ for a system governed by the Langevin dynamics follows the Fokker-Planck equation and can be described as:

$$\frac{\partial P(\boldsymbol{x},t)}{\partial t} = \frac{1}{\beta}\nabla\left[P(\boldsymbol{x},t)\nabla U(\boldsymbol{x},t) + kT\nabla P(\boldsymbol{x},t)\right] \tag{3}$$

With $t \to \infty$, the stationary distribution ($\partial P(\boldsymbol{x},t)/\partial t = 0$) is $P_\infty(\boldsymbol{x}) \propto exp(\frac{-U(\boldsymbol{x})}{kT})$. This indicates a lower interaction potential of particles always has a higher distribution probability in a Langevin dynamics system. Additionally, with temperature $T$ approaches zero, the probability distribution of the ground state significantly increases.

### C. Max-cut problem

Max-cut problem is a foundational NP-hard COP that facilitates the development of numerous real-world applications such as social networks, VLSI layout design, and machine learning. Given a graph $G = (V, W)$ with vertices $V$ and weighted edges $W$, the goal of the max-cut problem is to partition the vertices into two subsets $(S, \bar{S}) \in V$ such that the sum of weights of the edges crossing the two sets is maximized, as is shown in Fig. 1(c). The objective function of a max-cut problem can be expressed as Eqn. (4) [9]:

$$\mathcal{F}(G(V,W))_{max} = max\frac{1}{4}(\sum_{i,j} w_{ij} - \sum_{i,j} w_{ij}s_is_j) \tag{4}$$

Here, $w_{ij}$ is the weight value associated with edges $W$, and $s_i$, $s_j$ denote vertices in the two disjoint subsets $(S, \bar{S})$. The first term represents the summation of weighted edges in the graph, which is constant. Therefore, the max-cut problem can be exactly mapped to the Ising model without the external magnetic field, and the solution to the max-cut problem is the ground state of the Ising model. Hence, we utilize the max-cut problem as the benchmark to assess the performance of the annealing algorithm in solving the COPs.

## III. Annealing Algorithm Based on Langevin Dynamics

On the BRIM substrate, each spin collects the coupling current from connected spins via couplers. The cumulative current either charges or discharges the spin capacitor to maintain or flip the spins. Thus, random noise current can be injected along the current summation path to perturbate individual spin, leading to the spin output as:

$$\frac{dV_i}{dt} = \frac{1}{C}\left(\sum_{j\neq i} J_{ij}V_j + i_n\right) \tag{5}$$

According to BRIM, $C$ represents the capacitance value. $J_{ij} = sign(J_{ij})/R_{ij}$ is the coupling coefficient. Here, $V_i$ denotes the spin under analysis, and $V_j$ is the spin connected to it. $i_n = \kappa \cdot I_{\xi_i}$ is the random noise we want to inject, where $\kappa$ is the noise coefficient and $I_{\xi_i}$ is the white noise current. It is important to note that to effectively perturbate a spin within various sizes of Ising machines, the magnitude of noise current must be adjusted by the problem's scale. Specifically, it should be proportionate to the total coupled current for each spin. Therefore, we define the probability distribution of $I_{\xi_i}$ as Gaussian distribution $I_{\xi_i} = |\sum J_{ij}V_j| \cdot \hat{I}_{\xi_i}$ and $\hat{I}_{\xi_i} \sim N\left(0, \sigma^2\right)$ so that the noise current is comparable with the coupling current. Thus we rewrite the spin output as:

$$\frac{dV_i}{dt} = \frac{1}{C}\left(\sum_{j\neq i} J_{ij}V_j + \kappa\left|\sum_{j\neq i} J_{ij}V_j\right|\hat{I}_{\xi_i}\right) \tag{6}$$

It can be inferred that if the magnitude of white noise current is sufficiently large, for instance, $\left|\kappa\hat{I}_{\xi_i}\right| > 1$, the system will be dominant by random noise rather than Ising Hamiltonian, thereby causing the system to deviate from the equilibrium point such as local minima and engage in a fluctuation. The fluctuation can be used for exploring solution space and searching for all possible solutions. Conversely, when $\left|\kappa\hat{I}_{\xi_i}\right| \leq 1$, the Ising Hamiltonian becomes dominant and we can define a Lyapunov function $\mathcal{L}$ and find its derivative as shown in Eqn. (7).

$$\mathcal{L}(\boldsymbol{V}) = -\sum_i\left(\sum_{j\neq i} J_{ij}\int V_j\right), \quad \boldsymbol{V}^\top = (V_1, V_2\cdots V_N)^\top$$

$$\frac{d\mathcal{L}}{dt} = \sum_n \frac{\partial\mathcal{L}(V_n)}{\partial V_n}\frac{\partial V_n}{\partial t} = -\frac{(1+\kappa\hat{I}_{\xi_i})}{C}\left(\sum_{j\neq i} J_{ij}V_j\right)^2 \tag{7}$$

Where $\left|\kappa\hat{I}_{\xi_i}\right| \leq 1$ and $N$ is the total number of spins in the Ising model. It can be seen that the Lyapunov function is non-positive, indicating that the system will inevitably evolve towards decreasing the scaled Ising Hamiltonian as shown in Eqn. (8) and eventually reach equilibrium points.

$$H(\boldsymbol{V}) = -\frac{1}{C}\left(\sum_{j\neq i} J_{ij}V_iV_j\right) + \frac{\kappa}{C}P_{\xi_i} \tag{8}$$

Here $P_{\xi_i}$ is the power of random noise perturbating the system. Note that the system becomes governed by Langevin dynamics that includes an interaction potential and white noise based on Eqn. (2) and Eqn. (8). Following the Fokker-Planck equation, we can obtain the evolution of the joint probability density function as written in Eqn. (9). The lower the Ising Hamiltonian, the more possible the system will eventually converge to the corresponding state. With $\kappa$ being lowered to zero, the probability of the system reaching the global minimum of the Ising Hamiltonian is gradually enhanced.

$$P_\infty(\boldsymbol{V}) = \mathcal{N}\cdot exp\left(\frac{-H(\boldsymbol{V})C^2}{\kappa^2}\right), \quad \mathcal{N}\ is\ constant \tag{9}$$

Hence, our new annealing schedule involves three steps:

1). **Perturbation:** At the beginning of the annealing, set $\kappa > 1$ to thoroughly perturbate the system, facilitating exploration across the solution space.

2). **Annealing:** Gradually decrease $\kappa$ throughout annealing, thereby increasing the likelihood of the system transitioning into the ground state. It is noteworthy that the duration of annealing varies depending on the problem's scale.

3). **Equilibrium:** Cease noise injection and allow the system to converge towards the equilibrium state.

Algorithm 1 outlines the annealing schedule.

**Algorithm 1** Langevin dynamics annealing schedule

**Require:** $J$: coupling matrix, $R$: resistance, $C$: capacitance, $\sigma_0$: initial spin, $t_{stop}$: annealing time, $\Delta t$: time step, $I_{N0}$: initial noise amplitude, $\Delta I_N$: noise step

1: **function** ANNEAL($J$, $R$, $C$, $\sigma_0$, $t_{stop}$, $\Delta t$, $I_{N0}$, $\Delta I_N$)
2: $\sigma \leftarrow \sigma_0$
3: $I_N \leftarrow I_{N0}$
4: $steps \leftarrow t_{stop}/\Delta t$
5: $\Delta I_N \leftarrow (I_{N0} - 0)/steps$
6: **for** $k = 0$ **to** $steps - 1$ **do**
7: $I \leftarrow JV/(RC) + I_N$ ▷ Coupling current + noise
8: $V \leftarrow V + \Delta t \cdot I/C$ ▷ Update spin
9: $I_N \leftarrow I_N - \Delta I_N$ ▷ Update noise amplitude
10: **end for**
11: **return** $\sigma$
12: **end function**

## IV. Hardware implementation

Applying the preceding theory to hardware implementation, we will have a detailed discussion on the hardware implementation of the proposed annealing algorithm in this section, including system-level and circuit-level design.

### A. System architecture design

*1) System overview:* As illustrated in Fig. 2, we implemented a comprehensive $N$-spin Ising machine based on BRIM to validate the performance of our proposed annealing algorithm. The system architecture comprises an off-chip processor, on-chip central control logic, a Langevin dynamics annealing module, on-chip SRAM, and the BRIM itself. Note that the on/off-chip control logic and memory primarily handle problem mapping onto the BRIM system, but do not directly participate in computations, aside from controlling the annealing module.

*2) Communication and control:* The off-chip processor is responsible for ingesting real-world problem data and mapping it onto the Ising machine chip. With a certain number of iterations completed, the processor retrieves the solutions from the on-chip memory and evaluates the Ising Hamiltonian. The central control logic manages the entire workflow, including initializing the spins, configuring the coupling units, executing the annealing schedule, and reading out the final solutions, all operating in a pipeline. To accelerate the programming of the coupling units, the $N \times N$ coupling array is programmed in parallel by $N$ DACs, which work in a column-wise fashion, sequentially programming each column of coupling units via a counter. The on-chip SRAM is configured to store the initial and final spins, $M$-bit coupling coefficients, and coupling polarity. Communication between the off-chip processor and the on-chip central control logic is achieved via SPI and high-speed IOs.

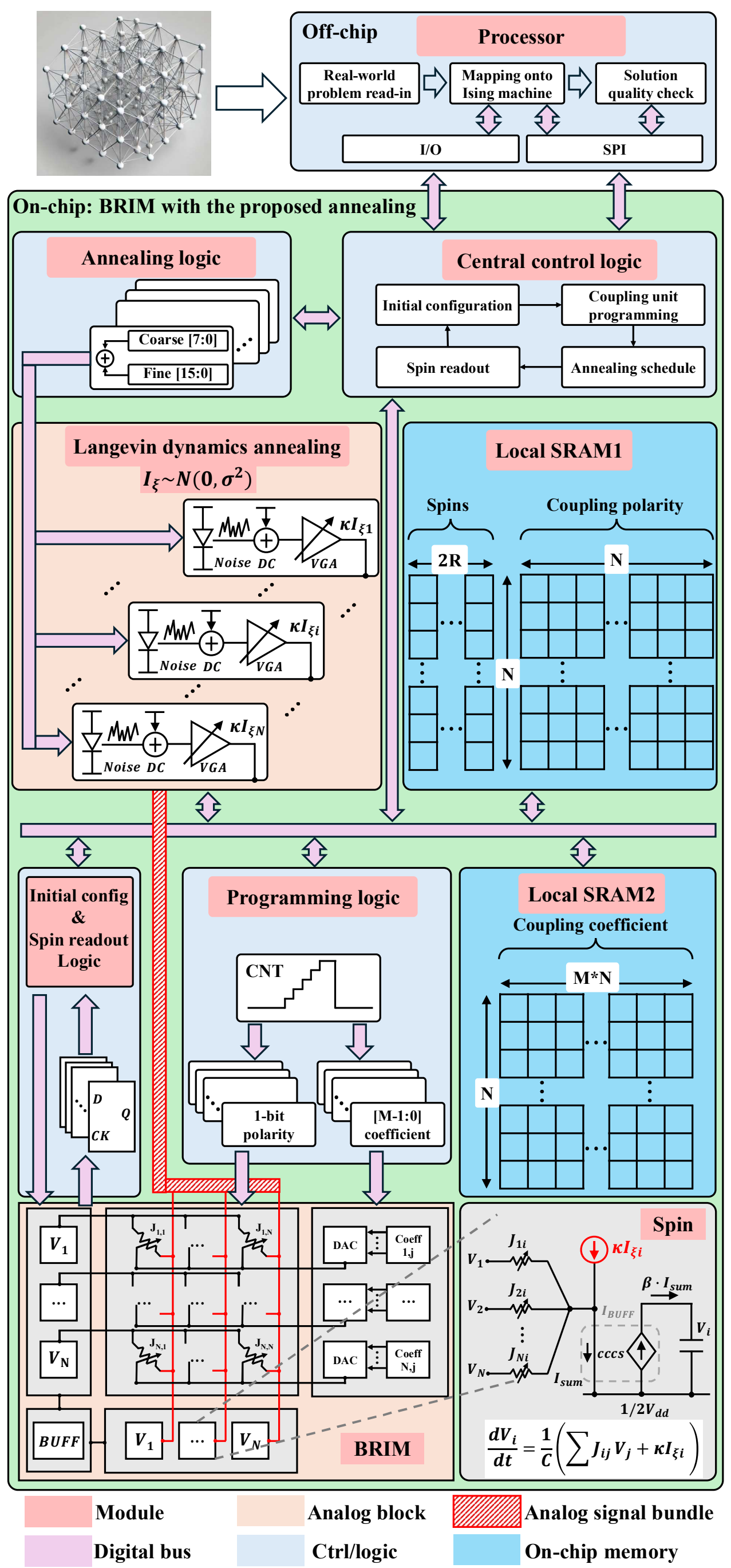

Fig. 2. System architecture of BRIM with annealing algorithm based on Langevin dynamics.

*3) Langevin dynamics:* The Langevin dynamics annealing module includes a noise generator, a DC adjuster, and a variable gain amplifier (VGA). It operates with 8-bit coarse static tuning for determining the initial noise amplitude, and 16-bit fine dynamic tuning to control the annealing schedule. Both tuning mechanisms are managed by the annealing logic, operating within a designated annealing time window to achieve the algorithm 1. During system evolution, random noise currents are injected directly into the current summation ports of the spins. Leveraging the low input impedance characteristic of the input buffer of spin, the injected noise current does not affect the input voltage of the spins, ensuring it does not interfere with the coupling current from other spins.

### B. Circuit level implementation

With the system design illustrated earlier, we now delve into the circuit implementation of the proposed annealing algorithm, focusing specifically on the design of the noise generator, DC adjuster, and VGA as depicted in Fig. 3.

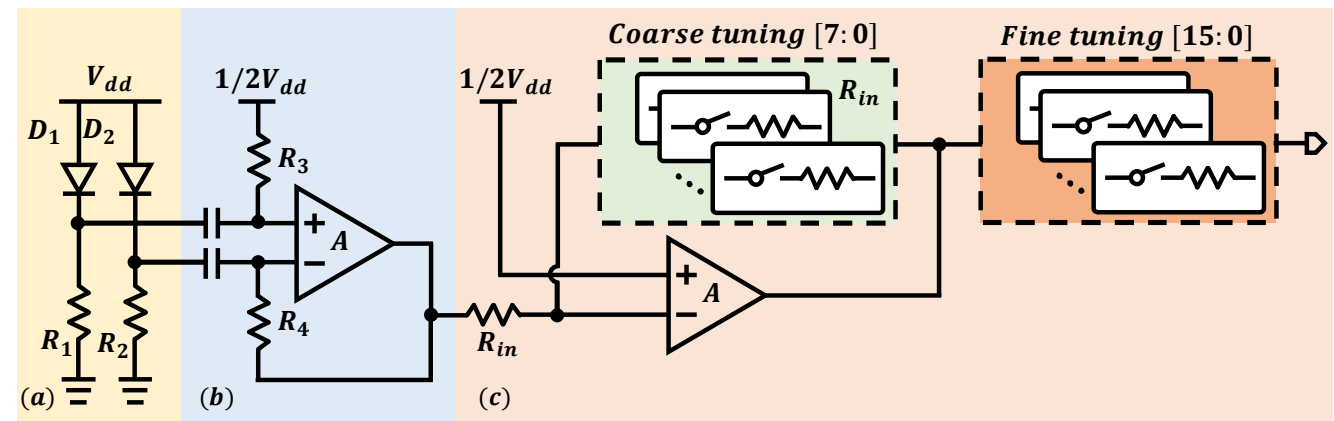

Fig. 3. Circuit level implementation of noise generator for annealing algorithm based on Langevin dynamics including (a) noise source, (b) DC adjuster, and (c) variable gain amplifier.

Thermal noise in electronic components, such as diodes and resistors, has been widely studied due to its inherent randomness, which makes it a valuable source for generating random perturbations. By amplifying this noise to adjustable levels, Langevin dynamics annealing can be achieved, as illustrated in Fig. 3 (a).

In the context of the BRIM, the *DC* voltage at the input port of the current buffer is set to $1/2V_{dd}$, establishing a virtual ground. This configuration enables the mapping of the digital 1 and 0 to $+1$ and $-1$, respectively, in the Ising model. Consequently, it is notable that the DC level of the noise signal should also be set to $1/2V_{dd}$ to prevent the continuous injection of a *DC* current into spins, which could otherwise cause the spin to remain fixed at one of the power rails. To address this, a DC adjustment module is incorporated between the noise generator and the variable gain amplifier as shown in Fig. 3 (b), ensuring that the noise signal is appropriately biased before it interacts with the spin system.

The variable gain amplifier is designed with two key adjustment mechanisms, as detailed in Fig. 3 (c). The first mechanism is an 8-bit coarse tuning, which modifies the amplifier's gain by adjusting the feedback resistor values. This step is configured before computation and is dependent on the problem size to set the initial noise amplitude in the annealing schedule. The second mechanism is a 16-bit fine-tuning process, managed by a finite state machine (FSM) during each iteration of the computation. This fine-tuning gradually decreases the noise amplitude, guiding the system towards convergence at the ground state.

## V. Performance Evaluation and Analysis

In this section, we did simulations using the proposed hardware annealing algorithm on the BRIM substrate to evaluate its performance. Given the direct mapping between the MAX-CUT problem and the Ising formula, we used MAX-CUT problems as a testbench in our simulations. The evaluation was performed on two different levels:

1). **Chip-level:** A 50-spin Ising machine was developed using commercial 65nm CMOS technology, with analog circuit blocks fully implemented and digital blocks synthesized. Simulations were performed on the Cadence Virtuoso platform with the Spectre as the analog domain simulator and the analog mixed-signal (AMS) tool as the chip-level simulator. These simulations enabled an in-depth analysis of the chip's performance, including solution quality, system's transient response, and detailed insights into layout and area breakdown.

2). **Behavior-level:** In order to evaluate the scalability of the annealing algorithm for larger COPs, behavioral-level simulations were conducted on a larger system. The performance was compared against the spin-flip, using the same BRIM substrate, in terms of proximity to the best-known solution (BKS) and TTS. Additionally, we explored the effects of device variation on the performance of the annealing algorithm.

## *A. Chip-level performance*

Without loss of generality, 10 randomly generated all-to-all connected MAX-CUT problems with binary weights were mapped onto the Ising machine. Each problem was run 50 times with different initial spin configurations, maintaining a fixed simulation time of 100 ns. The top 20 results are summarized in Fig. 4, with the analysis focusing on two key performance metrics: (1) the probability of reaching the BKS, where a higher probability is preferable, and (2) the average distance from the BKS, where a lower value is desirable.

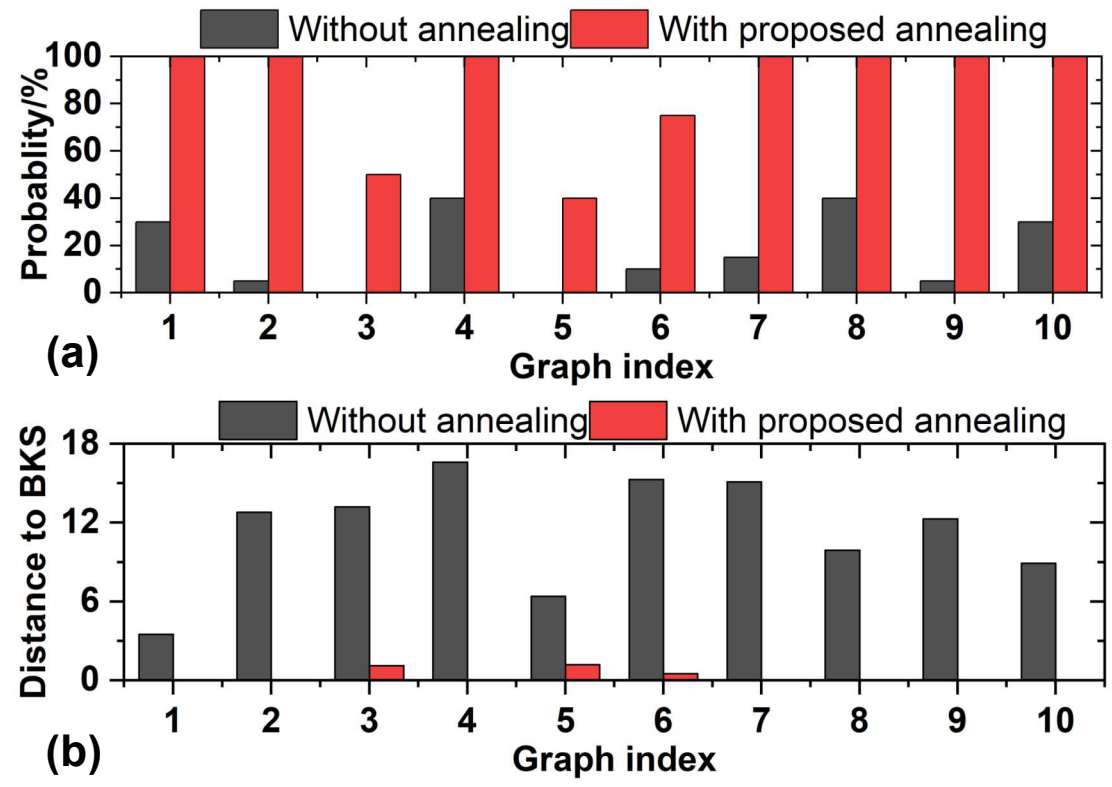

Fig. 4. Chip-level performance of (a) probability to reach ground state and (b) distance from the BKS.

Without hardware annealing, the system's probability of reaching the BKS was 17.5%, with an average distance from the BKS of 11.4. However, with the proposed hardware annealing algorithm, the system reached the ground state with a significantly improved probability of 86.5%, nearly 5 times higher than without annealing. Additionally, the average distance from the BKS dropped to 0.28, representing a 97.5% improvement.

We further analyzed the transient response of the proposed hardware annealing approach. Fig. 5 illustrates the transient waveform from a representative simulation case for graph 1. To thoroughly perturb the system, we initialized the noise amplitude at $90\mu A$, ensuring that $\kappa > 1$ for every spin. During the annealing process, the system actively explores various energy states, quickly converging towards a low-energy state. Eventually, the system reaches the ground state within the prescribed annealing schedule, as the energy landscape navigation indicated by the blue curve.

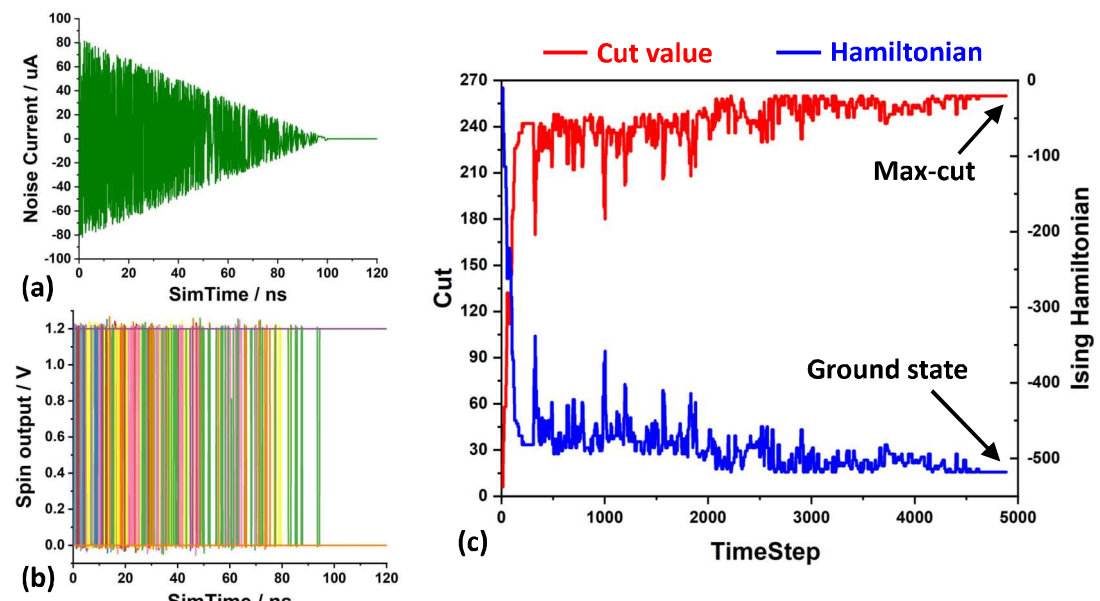

Fig. 5. Transient waveform of (a) Langevin noise current and (b) spin output (c) Cut value and Ising Hamiltonian versus the simulation time step.

Fig. 6 illustrates the chip layout and area breakdown for the proposed hardware annealing system. The noise generator is positioned adjacent to the current summation ports of the spins, allowing it to directly inject noise current into them. This can remove the need for control logic operations such as reading and flipping the spins, thereby reducing interconnection overhead in the layout.

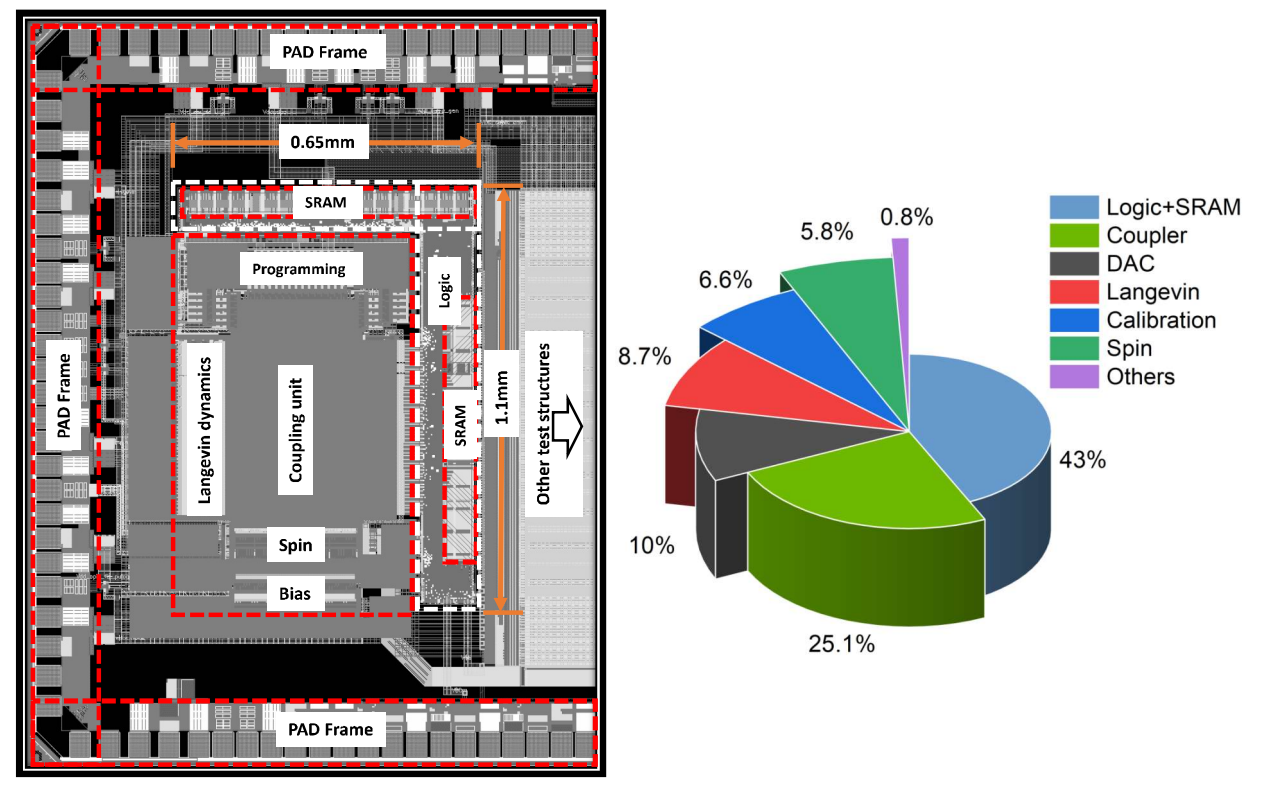

Fig. 6. Layout (left) and area breakdown (right) of BRIM with the proposed hardware annealing.

## *B. Behavior-level performance*

We benchmark the performance of the hardware annealing based on Langevin dynamics (**LD**) in the behavioral model of a 2000-spin BRIM by solving G-set max-cut problems [11]. We then compare the results with those obtained using spin-flip annealing (**SF**) on the same BRIM substrate [3]. Table I shows the distance from the BKS across 30 problems, , with both methods using an annealing time of $2.2\mu s$, consistent with the conditions in [3]. All the solutions are the best result from 50 different runs. The average distance from the BKS using the LD method is 1.8, while the SF method yields an average distance of 2.6. In particular, we analyzed the probability distribution of cut values and annealing process for G022, commonly used as benchmarks for Ising machines [13], as depicted in Fig. 7 (a). We further conducted a more focused comparison of solution quality between LD and SF under a constrained annealing time. Fig. 7 (b) shows the average distance from the BKS over 500 runs across the 30 reported problems at different annealing times. The results show that LD can reduce the TTS by 50% with achieving the same solution quality. It is important to note that the behavior simulation does not include the time spent on spin updates during SF annealing, which may underestimate the total time required by SF. As a result, LD demonstrates superior performance in terms of overall TTS.

TABLE I
COMPARISON OF DISTANCES TO BKS FOR **LD** AND **SF**.

| **Graph** | **This work** | | **Ref [3]** | **Graph** | **This work** | | **Ref [3]** |
|---|---|---|---|---|---|---|---|
| | **LD** | **SF** | **SF** | | **LD** | **SF** | **SF** |
| G001 | 0 | 0 | 0 | G016 | 5 | 9 | 6 |
| G002 | 0 | 0 | 0 | G017 | 3 | 7 | 6 |
| G003 | 0 | 0 | 0 | G018 | 4 | 4 | 4 |
| G004 | 0 | 0 | 0 | G019 | 0 | 1 | 3 |
| G005 | 0 | 0 | 0 | G020 | 0 | 0 | 0 |
| G006 | 0 | 0 | 0 | G021 | 2 | 2 | 3 |
| G007 | 0 | 0 | 0 | G043 | 0 | 3 | 0 |
| G008 | 0 | 0 | 0 | G044 | 0 | 4 | 0 |
| G009 | 1 | 0 | 1 | G045 | 0 | 8 | 1 |
| G010 | 1 | 1 | 1 | G046 | 0 | 6 | 2 |
| G011 | 4 | 12 | 6 | G047 | 1 | 8 | 1 |
| G012 | 2 | 6 | 4 | G051 | 5 | 7 | 3 |
| G013 | 4 | 8 | 4 | G052 | 5 | 1 | 9 |
| G014 | 6 | 8 | 3 | G053 | 5 | 6 | 10 |
| G015 | 2 | 4 | 4 | G054 | 4 | 3 | 8 |

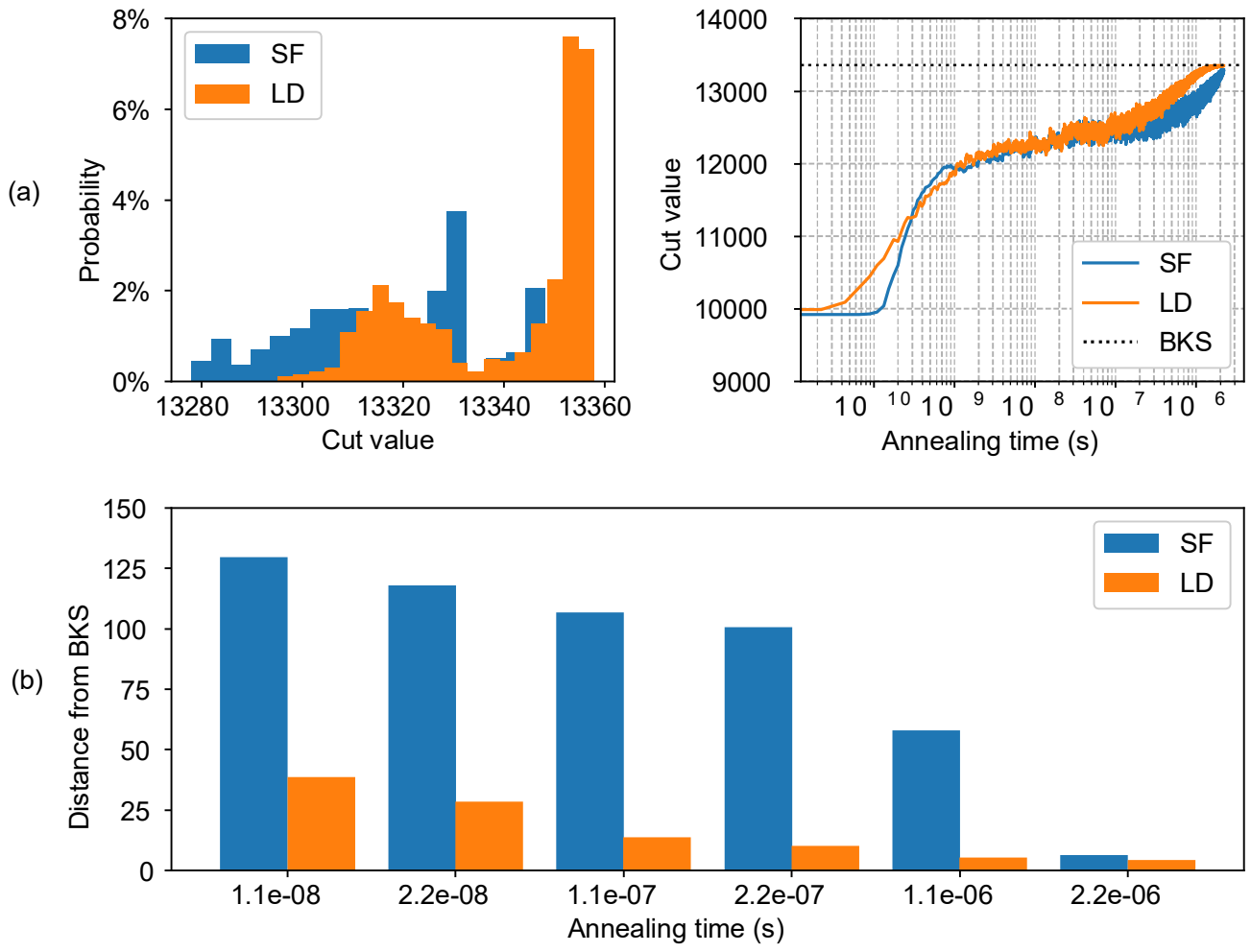

Fig. 7. (a) Probability distribution of cut values and annealing process for G022. (b) Average distance from the BKS across 30 G-set problems.

## C. Device variation

For the proposed noise generator based on analog devices, PVT variations should be taken into account. Thus, we applied 3%, 5%, 10%, and 20% variations to the noise current magnitude in behavioral simulations and compared the solution qualities within different annealing times for G022 and G039, as shown in Fig. 8. It can be seen that the proposed hardware annealing has a robust resistance to the PVT variation generated on the injected current.

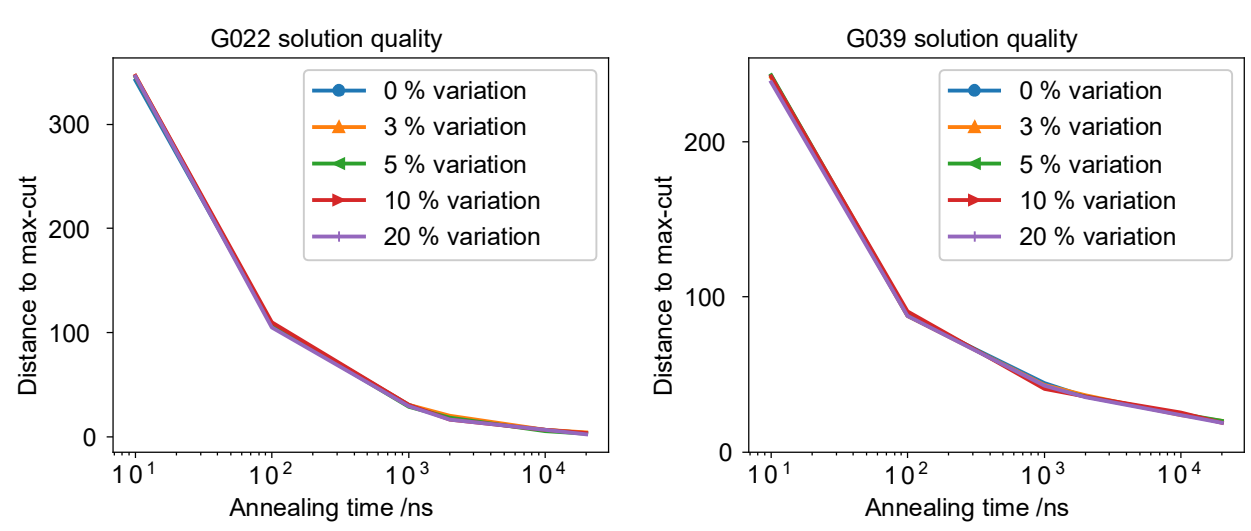

Fig. 8. Solution qualities versus variations and annealing time for G022/G039.

# VI. CONCLUSION

In conclusion, the proposed hardware annealing algorithm for Ising machines, based on Langevin dynamics, demonstrates promising advantages over alternative algorithms. Simulation results show that the algorithm effectively assists in escaping local minima with a short TTS. The results also suggest that the proposed algorithm holds the potential for enhancing the performance of Ising machines.